\documentclass[eqsecnum,twocolumn,showpacs,preprintnumbers,amssymb,aps]{revtex4}
\usepackage{graphicx}
\usepackage{slashbox}
\usepackage{dcolumn}
\usepackage{bm}
\usepackage{latexsym,epsfig}

\begin{document}

\preprint{\today}

\title{Nuclear quadrupole moment of $^{43}$Ca and hyperfine structure studies of its singly charged ion}
\vspace{0.5cm}

\author{B. K. Sahoo \protect \footnote[1]{E-mail: B.K.Sahoo@rug.nl}}
\affiliation{KVI, University of Groningen, NL-9747 AA Groningen, The Netherlands}
\date{\today}
\vskip1.0cm

\begin{abstract}
\noindent
By combining our theoretical calculation and recently measured electric
quadrupole hyperfine structure constant of the $3d \ ^2D_{5/2}$ state in the
singly ionized $^{43}$Ca, we determine its nuclear quadrupole moment to one 
percent accuracy. The obtained result, $-0.0444(6)b$, is about ten percent
improvement over the considered standard value. We have employed the
relativistic coupled-cluster theory at single and double excitations level
to calculate the electronic wave functions. The accuracy of these wave
functions are estimated by comparing our calculated magnetic dipole 
hyperfine constants with their corresponding available experimental results of many low-lying
states. We also present hyperfine structure constants for other higher excited 
states where experimental results are not reported. Role of the Breit
interaction has been investigated in these properties.
\end{abstract}


\maketitle

\section{Introduction}
Advanced modern techniques of laser cooling and trapping have enabled us to
carry out precision measurements of hyperfine structure constants in atomic
systems \cite{ghosh,demtroeder}. Theoretical studies of these quantities 
require accurate many-body methods, inclusion of relativistic effects and
knowledge of nuclear moments \cite{karshenboim,boucard,schwartz,das}.
Precise measurements of nuclear moments are difficult, especially the 
quadrupole and octupole moments. The prominent
examples of techniques to measure them are NMR, atomic beams, optical pumping,
recoil methods etc. \cite{kalish,keim}. However, their absolute results
are of great interest for the nuclear physicists to be able to test different
nuclear models \cite{neyens,arnold}. The quadrupole moment of the stable isotope
$^{43}$Ca is also of particular interest in the evaluation of the nuclear
magnetic resonance measurements in the biological systems \cite{wong,forsen}.
Investigating properties of nuclei in the region of magic numbers
are challenging because the valence nucleons can be strongly affected
by the close shell configuration. $^{43}$Ca has mass number between the
double magic numbers $^{40}$Ca and $^{48}$Ca suggesting peculiar nuclear charge
distribution. It is plausible to obtain correct electronic wave functions,
hence their properties, in the single valence atomic systems using the recently
developed highly
potential methods like all order relativistic coupled-cluster (RCC) theory
\cite{das,sur,sahoo1,sahoo2}. Singly ionized $^{43}$Ca ($^{43}$Ca$^+$) is of 
medium size system and single and double excitations approximation RCC approach
(CCSD method) is capable enough in such cases \cite{sur,sahoo1,sahoo2} 
to account electron correlation effects accurately. It has been observed that 
electron
correlation effects exhibit spectacular behavior in the studies of the
magnetic dipole hyperfine structure constant in the $^2D_{5/2}$ states of
the singly ionized alkaline-earth metal atoms. The CCSD method with 
contributions from leading order triple excitations (CCSD(T) method) is
able to consider them sufficiently for which it can produce results matching
with the precisely measured values \cite{sahoo2,sahoo3,sahoo4}.
In this work, we have used this method to evaluate the electronic matrix 
 elements due to the hyperfine interaction operators.

It would also be imperative to mention here that $^{43}$Ca$^+$ is an interesting
candidate for quantum computation \cite{roos1} and optical frequency standard
\cite{ito,benhelm1}. Accurate values of hyperfine structure constants are useful
in estimating shifts in the energy levels due to the stray electromagnetic 
fields during the experimental set-up \cite{itano1,itano2}. Again, theoretical 
estimations of these quantities are used to test the correct behavior of wave
functions in the nuclear region \cite{sahoo5,wansbeek}. Nuclear magnetic moment
of $^{43}$Ca is known within sub-one percent accuracy \cite{lutz}, but its
reported nuclear quadrupole moment values vary from $-0.0408(8)b$ to 
$-0.065(20)b$ \cite{sundholm,bergmann,grundevik,salomonson,silverans}. Using 
the accurate matrix element of the electric quadrupole hyperfine interaction
operator and precisely measured \cite{benhelm2} electric quadrupole hyperfine 
structure constant of the $3d \ ^2D_{5/2}$ state in $^{43}$Ca$^+$, we 
determine its nuclear quadrupole moment. Accuracy of the theoretical 
calculations are estimated from the analysis of the correct behavior of the
wave functions which are able to reproduce the magnetic dipole hyperfine
structure constants in a few low-lying states where experimental results are
available. We also present hyperfine structure constants of other higher 
excited states for further testimony of our results using future experiments.

\section{Theory and Method of Calculations}
The detail theory about the hyperfine structures are given earlier in a
classic paper by C. Swartz \cite{schwartz}. Here, we have mentioned
only these formulae in explicit form. The relativistic hyperfine interaction 
Hamiltonian is given by
\begin{equation}
H_{hfs} = \sum_l {\bf{M}}^{(l)}\cdot {\bf{T}}^{(l)}
\label{eqn1}
\end{equation}
where ${\bf{M}}^{(l)}$ and ${\bf{T}}^{(l)}$ are spherical tensor operators of
rank $l$. In the first-order perturbation theory, hyperfine interaction energy 
$W_F$ of the hyperfine
state $|F;I,J \rangle$ of angular momentum $F=I+J$ with $I$ and $J$ being the
nuclear spin and electronic angular momentum of the associated fine structure 
state $|J,M_J\rangle$, respectively, after neglecting terms beyond $l=2$
is given by
\begin{equation}
W_{F}= \frac{1}{2} A_{hfs} K + B_{hfs} \frac{\frac{3}{2}K(K+1)-2I(I+1)J(J+1)}{2I(2I-1)2J(2J-1)},
\label{eqn2}
\end{equation}
where $K=F(F+1)-I(I+1)-J(J+1)$, $A_{hfs}$ is the magnetic dipole structure 
constant for $l=1$ and $B_{hfs}$ is the electric quadrupole structure constant
for $l=2$. These constants are defined as
\begin{eqnarray}
A_{hfs} = \mu_N g_I \ \frac {\langle J || \textbf{T}^{(1)}||J\rangle}{\sqrt{J(J+1)(2J+1)}}, 
\label{eqn3}
\end{eqnarray}
and
\begin{eqnarray}
B_{hfs} = Q_{nuc} \left \{ \frac {8J(2J-1)}{(2J+1)(2J+2)(2J+3)} \right \}^{1/2} \langle J|| \textbf{T}^{(2)}||J\rangle, \ \
\label{eqn4}
\end{eqnarray}
where we have used atomic unit (au). In the above expressions, $\mu_N$ is the
Bohr magneton and we use $g_I= [\frac {\mu_I}{I}]$ with $\mu_I$ and $I(=7/2)$ 
are the nuclear magnetic dipole moment and spin, respectively, as $-0.37646943$ from the measurement 
\cite{lutz} to evaluate $A_{hfs}$. Since nuclear quadrupole moment, $Q_{nuc}$, 
is not known accurately, hence we calculate
\begin{eqnarray}
\frac{B_{hfs}}{Q_{nuc}} = \left \{ \frac {8J(2J-1)}{(2J+1)(2J+2)(2J+3)} \right \}^{1/2} \langle J|| \textbf{T}^{(2)}||J\rangle. \ \
\label{eqn5}
\end{eqnarray}

The reduced matrix elements of the electronic spherical operators, $\textbf{T}^{(l)}=\sum t^{(l)}$, in terms of single orbitals are given by
\begin{eqnarray}
\langle \kappa_f || t^{(1)} || \kappa_i \rangle = - (\kappa_f+\kappa_i) \langle -\kappa_f || C^{(1)} || \kappa_i \rangle \nonumber \\ \int_0^{\infty} dr \frac{(P_fQ_i+Q_fP_i)}{r^2}
\label{eqn6}
\end{eqnarray}
and
\begin{eqnarray}
\langle \kappa_f || t^{(2)} || \kappa_i \rangle = - \langle \kappa_f || C^{(2)} || \kappa_i \rangle \int_0^{\infty} dr \frac{(P_fP_i+Q_fQ_i)}{r^3}, \ \ \ \
\label{eqn7}
\end{eqnarray}
where $i$ and $f$ represent initial and final orbitals, respectively. The reduced matrix elements of the spherical tensors ($C^{(l)}$) are given by
\begin{eqnarray}
\langle \kappa_f || C^{(l)} || \kappa_i \rangle = (-1)^{j_f+1/2} \sqrt{(2j_f+1)(2j_i+1)} \nonumber \\ \left ( \matrix { j_f & l & j_i \cr 1/2 & 0 & -1/2 \cr } \right ) \pi(l_f,l,l_i)
\label{eqn8}
\end{eqnarray}
with the angular momentum selection rule $\pi(l_f,l,l_i)=1$ when $l_f+l+l_i=$even for the orbital angular momenta $l_f$ and $l_i$, otherwise zero. 

The primary objective of this work is to calculate the above electronic matrix elements 
of the hyperfine interaction operators. It is obvious from the single particle
expressions that these matrix elements have strong overlap with the nucleus.
As a starting point, we consider kinetically balanced Gaussian type of orbitals (GTOs) which seem
to be an ideal choice for obtaining correct behavior of wave functions in the 
nuclear region \cite{ishikawa,mohanty} to calculate mean-field wave functions $|\Phi_0 \rangle$
 of the closed-shell configuration in the Dirac(Hartree)-Fock (DF) approach. 
To calculate the atomic state function (ASF) of single valence with 
closed-shell configurations, we express it in the RCC ansatz as 
\begin{eqnarray}
|\Psi_v \rangle = e^T \{1+S_v\} |\Phi_v \rangle ,
\label{eqn9}
\end{eqnarray} 
where $|\Phi_v \rangle$ is the new reference state which is defined as 
$|\Phi_v \rangle=a_v^{\dagger}|\Phi_0\rangle$ and will give the DF result for
the above open-shell configuration. In the above expression, we call $T$ and 
$S_v$ as the closed and open shell core and core with valence electron 
excitation operators, respectively, which in the second quantization notation
in the CCSD approximation are given by 
\begin{eqnarray}
T=T_1+T_2 = \sum_{a,p}a_p^+a_a t_a^p + \frac {1}{4}\sum_{ab,pq}a_p^+a_q^+a_ba_a t^{pq}_{ab}, \ \ \ \label{eqn4} \\
S_v=S_{1v}+S_{2v}= \sum_{p \ne v}a_p^+a_v s_v^p + \frac {1}{2}\sum_{b,pq}a_p^+a_q^+a_ba_v s^{pq}_{vb} , \ \
\label{eqn10}
\end{eqnarray}
where the ($a,b,c..$), ($p,q,r..$) and ($v$) subscripts of the second quantized 
operators represent core, particle (virtual) and valence orbitals, respectively. The $t$ and $s_v$  coefficients are the corresponding excitation amplitudes 
which are determined using the following equations
\begin{eqnarray}
\langle \Phi^L |\{\widehat{H_Ne^T}\}|\Phi_0 \rangle = 0  \ \ \ \ \ \ \ \ \ \ \ \ \ \ \ \ \ \ \ \ \ \ \ \ \ \ \ \label{eqn11} \\
\langle \Phi_v^L|\{\widehat{H_Ne^T}\}S_v|\Phi_v\rangle = - \langle \Phi_v^L|\{\widehat{H_Ne^T}\}|\Phi_v\rangle \ \ \ \ \ \ \ \ \ \ \ \nonumber \\ \ \ \ \ \ \ \ \ \ \ \ \ \ \ \ + \langle \Phi_v^L|S_v|\Phi_v\rangle \Delta E_v \ \delta_{L,0} ,
\label{eqn12}
\end{eqnarray}
with the superscript $L (=1,2)$ representing the single and double excited 
states from the corresponding reference states and the wide-hat symbol denotes
the linked terms. $\Delta E_v$ is the  corresponding valence electron affinity 
energy which is evaluated by
\begin{eqnarray}
 \Delta E_v = \langle \Phi_v|\{\widehat{H_N e^T}\} \{1+S_v\} |\Phi_v\rangle .
\label{eqn13}
\end{eqnarray}

In our CCSD(T) approach, we consider effects of the leading order triple
excitations through $\Delta E_v$ by constructing triple excitation operator 
\begin{equation}
S_{vbc}^{pqr}\ =\ \frac{\widehat{H_N T_2}+\widehat{H_N S_{v2}}}{\epsilon_b+\epsilon_c-\epsilon_q-\epsilon_r},
\label{eqn14}
\end{equation}
and contracting $S_{vbc}^{pqr}$ with the Hamiltonian to get contributions to the corresponding $\Delta E_v$, where $\epsilon_i$ is the DF energy of 
the electron in the $i^{th}$ orbital.

We consider the Dirac-Coulomb-Breit Hamiltonian in the above equations that is given by
\begin{eqnarray}
H  = c \vec \alpha \cdot \vec {\text{p}} + (\beta -1) c^2 + V_{nuc}(r) 
 + \frac {1} {r_{12}} - \frac {\vec \alpha_1 \cdot \vec \alpha_2 } {r_{12}} + \nonumber \\ 
\frac {1}{2} \left \{ \frac {\vec \alpha_1 \cdot \vec \alpha_2} {r_{12}}  - 
\frac {(\vec \alpha_1 \cdot \vec {\text{r}}_{12}) (\vec \alpha_2 \cdot \vec 
{\text{r}}_{12})} {r_{12}^3} \right \} ,
\label{eqn15}
\end{eqnarray}
where $c$ is the velocity of light, $\alpha$ and $\beta$ are the Dirac matrices
and $V_{nuc}(r)$ is the nuclear potential. We solve the wave functions due to
the above Hamiltonian in the DF and RCC methods self-consistently with the
tolerance size below $10^{-7}$ to obtain precise results.

\begin{table}
\caption{Results of $A_{hfs}$, $B_{hfs}/Q_{nuc}$ and $B_{hfs}$ of many states in $^{43}$Ca$^+$.}
\begin{ruledtabular}
\begin{center}
\begin{tabular}{lllll}
 &  \multicolumn{2}{c}{$A_{hfs}$} & $B_{hfs}/Q_{nuc}$ & $B_{hfs}$ \\
     \cline{2-3} &  Calc & Expt  & Calc & Expt \\
            &   (MHz) & (MHz) & (MHz$b^{-1}$) & (MHz) \\
\hline
               &            &  & & \\
4s $^2S_{1/2}$ & $-$806.387(1.0)$^a$ & $-$797.5(2.4)$^e$ &  \\
               & $-$805.348$^b$ & $-$805(2)$^f$ &  \\
               & $-$819$^c$ & $-$817(15)$^g$  &  \\
               & $-$794.7$^d$ &   & \\
               &         &             & \\
5s $^2S_{1/2}$ & $-$234.0(2.0)$^a$ & &  \\
               &         &             & \\
4p $^2P_{1/2}$ & $-$145.422(340)$^a$ & $-$158.0(3.3)$^e$ &  \\
                & $-$143.068$^b$ & $-$145.5(1.0)$^f$ &  \\
                & $-$148$^c$ & $-$142(8)$^h$ &  \\
                & $-$144.8$^d$ & 145.4(0.1)$^i$ &  \\
               &               &        & \\
5p $^2P_{1/2}$ & $-$49.651(521)$^a$ &  &  \\
               &               &        & \\
4p $^2P_{3/2}$ & $-$30.370(415)$^a$ & $-$29.7(1.6)$^e$ & 151.246(752)$^a$ & \\
            & $-$30.498$^b$ & $-$31.9(2)$^f$ & 151.798$^b$ & $-$6.7(1.4)$^f$ \\
                & $-$30.9$^c$ & $-$31.0(2)$^i$ & 155$^c$ & $-$6.9(1.7)$^i$\\
                & $-$29.3$^d$ &  & \\
               &            & & \\
5p $^2P_{3/2}$ & $-$10.335(112)$^a$ &  & 50.904(571)$^a$ \\
               &            & & \\
3d $^2D_{3/2}$ & $-$47.282(289)$^a$ & $-$48.3(1.6)$^h$ & 67.280(688)$^a$ & \\
            & $-$47.824$^b$ & $-$47.3(2)$^i$ & 68.067$^b$  & $-$3.7(1.9)$^i$\\
                & $-$52$^c$ & & 68$^c$  \\
                & $-$49.4$^d$ & &  \\
                & $-$47.27$^j$ & & 72.06$^j$ \\
               &          &     & \\
4d $^2D_{3/2}$ & $-$9.474(042)$^a$ &  & 17.442(134)$^a$ \\
               &          &     & \\
3d $^2D_{5/2}$ & $-$3.622(327)$^a$ & $-$3.8(6)$^i$ & 95.472(1.276)$^a$ & $-$3.9(6.0)$^i$ \\
               & $-$3.552$^b$ & $-$3.8931(2)$^k$& 100.208$^b$ & $-$4.241(4)$^k$\\
                & $-$5.2$^c$ & & 97$^c$  \\
                & $-$4.2$^d$ & &  \\ 
                & $-$4.84$^j$ & & 102.45$^j$  \\
               &          &     & \\
4d $^2D_{4/2}$ & $-$2.971(035)$^a$ &  & 24.769(213)$^a$ \\
\end{tabular}
\end{center}
\end{ruledtabular}
$^a$This work; $^b$\cite{yu}; $^c$\cite{martensson1}; $^d$\cite{martensson2}; $^e$\cite{goble}; $^f$\cite{silverans}; $^g$\cite{kelly}; $^h$\cite{kurth}; $^i$\cite{norterhauser}; $^j$\cite{itano2} and $^k$\cite{benhelm2}.
\label{tab1}
\end{table}

We evaluate expectation values due to the hyperfine interaction operators
using our RCC method by
\begin{eqnarray}
< O > &=& \frac {<\Psi_v | O | \Psi_v >} {<\Psi_v|\Psi_v>} \nonumber \\
&=& \frac {< \Phi_v |\{1+S_v^{\dagger}\} \overline{O} \{1 +S_v\} | \Phi_v>} {\{1+S_v^{\dagger}\} \overline{N}_0 \{1 +S_v\} } \nonumber \\
&=& \frac {< \Phi_v |\{1+S_{1v}^{\dagger}+S_{2v}^{\dagger}\} \overline{O} \{1 +S_{1v}+S_{2v}\} | \Phi_v>} {\{1+S_{1v}^{\dagger}+S_{2v}^{\dagger}\} \overline{N}_0 \{1 +S_{1v}+S_{2v}\} } , \ \ \ \ \
\label{eqn10}
\end{eqnarray}
where $O$ is any of the operator, $\overline{O}=(e^{T^{\dagger}} O e^T)$ and 
$\overline{N}_0 = e^{T^{\dagger}} e^T$. Generally, both $\overline{O}$ and 
$\overline{N}_0$ in the RCC approach are non-terminative series. However, we
use a special trick to obtain their almost all the leading order contributions
using the Wick's generalized theorem \cite{lindgren}. In this procedure, we
evaluate first effective one-body, two body terms etc. step by step and
at the end sandwich them between the $S_v$ and its conjugate operators.
This procedure has already been demonstrated in our earlier works \cite{sahoo2,sahoo5,sahoo6,sahoo7}.
We also explicitly present contributions from the normalization factors 
evaluating them in the following way
\begin{eqnarray}
Norm = \langle \Psi_v | O | \Psi_v \rangle \{ \frac {1}{1+N_v} - 1 \},
\label{eqn13}
\end{eqnarray}
where $N_v=\{1+S_{1v}^{\dagger}+S_{2v}^{\dagger}\} \overline{N}_0 \{1 +S_{1v}+S_{2v}\}$.

\section{Results and discussions}
Earlier, we have studied behavior of the electron correlation effects in the 
magnetic dipole hyperfine structure constants in the considered system using 
the CCSD(T) method and GTOs for few low-lying states \cite{sahoo6} and later 
only in the 3d $^2D_{5/2}$ state \cite{sahoo2}. Due to the limitations with 
the available computational resources at that time, we had restricted our
calculations by considering only up to f-symmetry (orbital quantum number 
$l=3$) orbitals. Again, we had used basis along with g-symmetry orbitals in the latter
to find out the peculiar behavior of the core-polarization effects in the
$^2D_{5/2}$ states of the alkaline earth-metal ions, however detail 
investigations in the accuracy of the wave functions for other low-lying states
were not carried out in that work. In the present work, we use larger basis
functions up to g-symmetry that produces the other properties \cite{sahoo8}
including electron affinity energies of many low-lying states matching 
with the experimental results.

In Table \ref{tab1}, we present our $A_{hfs}$ and $B_{hfs}/Q_{nuc}$ (or $B_{hfs}$) results
of many states along with the available theoretical values and experimental 
measurements. Estimated errors in our calculations are given inside the 
parenthesis. Considered sources of these errors are basically two folds:
(a) numerical calculations with limitations over finite size basis functions
within g-symmetry and (b) approximations at the level of excitations in the
RCC approach. In the first case, we have tested our DF results for a set
of basis functions to achieve consistent results and possible discrepancies
in these results are assumed as one of the sources of errors. Second, we
approximate our level of excitations at the singles and doubles as Ca$^+$
is assumed as not too heavy system. However, equations to determine these 
amplitudes, in principle, should couple with the higher excitations for 
accurate calculations. Although our leading order triple excitations take
care of most of these contributions, we estimate the contributions from the
higher excitations by studying differences of results between the CCSD and 
CCSD(T) methods and evaluating lower order diagrams that may arise through 
the neglected triple excitations in the RCC approach. This procedure may not 
be sufficient enough, however it explains their importance qualitatively.
We have scaled these contributions to account as the upper values of the
second source of errors.
\begin{table}
\caption{Breit contributions ($\Delta_{Br}$) to the $A_{hfs}$ and $B_{hfs}/Q_{nuc}$ in MHz and MHz$b^{-1}$, respectively.}
\begin{ruledtabular}
\begin{center}
\begin{tabular}{lcc}
State & \multicolumn{2}{c} {$\Delta_{Br}$} \\
& $A_{hfs}$ &  $B_{hfs}/Q_{nuc}$ \\
\hline
& & \\
4s $^2S_{1/2}$ & $-$0.697 & \\
5s $^2S_{1/2}$ & $-$0.214 & \\
4p $^2P_{1/2}$ & $-$0.115 &  \\
5p $^2P_{1/2}$ & $-$0.063 &  \\
4p $^2P_{3/2}$ & 0.048 & $-$0.242 \\
5p $^2P_{3/2}$ & 0.014 & $-$0.054 \\
3d $^2D_{3/2}$ & $-$0.102 & 0.150 \\
4d $^2D_{3/2}$ & 0.135 & $-$0.231\\
3d $^2D_{5/2}$ & $-$0.102 & 0.385 \\
4d $^2D_{5/2}$ & $-$0.035 & $-$0.298 \\
\end{tabular}
\end{center}
\end{ruledtabular}
\label{tab2}
\end{table}

 It is obvious that our $A_{hfs}$ results match quite well within the uncertainties of available experimental values giving an indication that our calculated
wave functions are accurate enough in the nuclear region. There are also
experimental results available for $B_{hfs}$ in many cases, but none of them
are accurate enough except the recent measured value in the 3d $^2D_{5/2}$
state \cite{benhelm2}. It is obvious from Table \ref{tab1} that there are quite
well agreement between different calculations of $B_{hfs}/Q_{nuc}$ results
at least in the 4p $^2P_{3/2}$ and 3d $^2D_{3/2}$ states, but there comes
large discrepancies between the results in the 3d $^2D_{5/2}$ state. Due to
consistency among the theoretical calculations in the 4p $^2P_{3/2}$ and
3d $^2D_{3/2}$ states, it would indeed be appropriate to combine these results
with the measured $B_{hfs}$ values of their corresponding states to determine
less accurately known $Q_{nuc}$ in $^{43}$Ca. In contrast, $B_{hfs}$ of the
3d $^2D_{5/2}$ state is measured quite precisely and hence it is necessary
to use its calculated $B_{hfs}/Q_{nuc}$ result to combine with its 
measured $B_{hfs}$ value to determine $Q_{nuc}$ in the considered system.
Therefore, we would like to investigate possible reasons of the discrepancies
among the theoretical methods which are employed in these calculations.
First, we investigate the
role of the Breit interaction from which we can realize the effect of higher
relativistic effects in the above properties which was not considered in
the previous works, then we proceed with describing differences in the
inclusion of various electron correlation effects through the employed 
theoretical methods.
\begin{table*}
\caption{RCC contributions to the $A_{hfs}$ calculations.}
\label{tab3}
\begin{tabular}{lcccccccccc}
\hline
\hline
RCC terms & 4s $^2S_{1/2}$ & 5s $^2S_{1/2}$ & 4p $^2P_{1/2}$& 5p $^2P_{1/2}$ & 4p $^2P_{3/2}$ & 5p $^2P_{3/2}$ & 3d $^2D_{3/2}$ & 4d $^2D_{3/2}$ & 3d $^2D_{5/2}$ & 4d $^2D_{5/2}$ \\
\hline
 & & & \\
 $O$ (DF) & $-$587.902 & $-$181.120 & $-$101.559 & $-$36.396 & $-$19.669 & $-$7.056 & $-$33.409 & $-$8.104 & $-$14.235 & $-$3.455 \\
$O - \overline{O}$ & $-$1.626 & 0.111 & 1.076 & 0.370 & 0.215 & 0.080 & $-$0.604 & $-$0.079 & $-$0.283  & $-$0.041 \\
$\overline{O} S_{1v} + cc $ & $-$103.321 & $-$21.064 & $-$21.032 & $-$5.887 & $-$4.089 & $-$1.159 & $-$8.373 & $-$0.244 & $-$3.554 & $-$0.100 \\
$\overline{O} S_{2v} + cc $ & $-$102.224 & $-$28.843 & $-$19.997 & $-$6.345 & $-$5.386 & $-$1.812 & $-$3.120 & $-$0.926 & 15.803 & 1.138 \\
$S_{1v}^{\dagger} \overline{O} S_{1v}$ & $-$4.527 & $-$0.613 & $-$1.113 & $-$0.245 & $-$0.217 & $-$0.049 & $-$0.571 & $-$0.031 & $-$0.241 & $-$0.013 \\
$S_{1v}^{\dagger} \overline{O} S_{2v} + cc $ & $-$7.338 & $-$0.873 & $-$1.689 & $-$0.313 & $-$0.374 & $-$0.048 & $-$0.260 & 0.115 & 0.932 & $-$0.112 \\
$S_{2v}^{\dagger} \overline{O} S_{2v} + cc $ & $-$9.615 & $-$3.117 & $-$1.108 &  $-$0.355 & $-$0.837 & $-$0.348 & $-$2.017 & $-$0.384 & $-$2.125 & $-$0.443 \\
 $Norm$ & 8.466 & 1.519 & 1.107 & 0.273 & 0.232 & 0.057 & 1.072 & 0.179 & 0.081 & 0.055 \\
\hline
\hline
\end{tabular}
\end{table*}

For high accuracy calculations, it may be necessary to find out contributions
from higher order relativistic corrections than the Coulomb interaction; 
occurs through the exchange of longitudinal photons. The next important 
contribution from frequency independent Breit interaction due to the transverse
photon \cite{breit} can be assumed as a bench mark test to estimate how big
the neglected relativistic effects would be. In Table \ref{tab2}, we present
contributions from the Breit interaction in the $A_{hfs}$ and $B_{hfs}/Q_{nuc}$
calculations for different states. As seen, these contributions in the
considered system are not large and it is larger in the ground state than in 
the excited states. It seems from this study that higher order relativistic
corrections like bound state QED effects are not important for high precision
calculations in the present system and the results mostly depend on the
electron correlation effects.
\begin{table*}
\caption{RCC contributions to the $B_{hfs}/Q_{nuc}$ calculations.}
\label{tab4}
\begin{tabular}{lcccccc}
\hline
\hline
RCC terms & 4p $^2P_{3/2}$ & 5p $^2P_{3/2}$ & 3d $^2D_{3/2}$ & 4d $^2D_{3/2}$ & 3d $^2D_{5/2}$ & 4d $^2D_{5/2}$ \\
\hline
 & & & \\
 $O$ (DF) & 96.976 & 34.789 & 55.354 & 13.354 & 78.466 & 18.939 \\
$O - \overline{O}$ & $-$0.869 & $-$0.307 & 3.208 & 0.493 & 4.456 & 0.680 \\
$\overline{O} S_{1v} + cc $ & 20.220 & 5.736 & 15.275 & 0.306 & 21.509 & 0.451 \\
$\overline{O} S_{2v} + cc $ & 32.885 & 10.217 & $-$5.066 & 2.836 & $-$6.843 & 4.067 \\
$S_{1v}^{\dagger} \overline{O} S_{1v}$ & 1.076 & 0.243 & 1.155 & 0.060 & 1.619 & 0.084\\ 
$S_{1v}^{\dagger} \overline{O} S_{2v} + cc $ & 1.770 & 0.286 & $-$1.586 & 0.456 & $-$2.225 & 0.632 \\ 
$S_{2v}^{\dagger} \overline{O} S_{2v} + cc $ & 0.341 & 0.220 & 0.621 & 0.308 & 0.844 & 0.432 \\
 $Norm$ & $-$1.153 & $-$0.280 & $-$1.681 & $-$0.371 & $-$2.354 & $-$0.516 \\
\hline
\hline
\end{tabular}
\end{table*}

We now discuss the differences between various works that account correlation
effects at various level of orders. As mentioned earlier, our previous and
present works are carried out with all order RCC method, but the main differences in the results are due to inclusion of orbitals from g-symmetry and Breit
interaction in this work. Both Yu et al. \cite{yu} and Martensson et al.
\cite{martensson1} have carried out their calculations using semi-empirical
feature many-body methods. They contain all order core-polarization effects,
however other correlation effects like Bruckner pair correlation effects 
\cite{lindgren} are taken up to certain orders. In the work of Yu et al., 
they have restricted orbitals in the evaluation of the hyperfine structure 
constants for individual state by selecting maximum contributing
angular momentum configurations. However, contributions from all these
orbitals are intrinsically accounted through the coupled equations in the
RCC method. In Tables \ref{tab3} and \ref{tab4}, we present contributions
from individual RCC terms to our $A_{hfs}$ and $B_{hfs}/Q_{nuc}$ results,
respectively. As seen our DF results match well with the Martensson et al., 
but there are differences between all order core-polarization contributions
between their work and ours. Our all order core-polarization effects are 
associated with the $S_{2v}$ RCC operator \cite{sahoo6,sahoo7}. The reason
of discrepancies could be due to the fact that all correlation effects are
coupled in the RCC method in contrast to the above semi-empirical features. 
There is also one non-relativistic theory with relativistic corrections
under random phase approximation (RPA) approach has been employed
\cite{martensson2} for calculating these hyperfine structure constants.
Recently, another calculation has been carried out using multi-configurational
Dirac-Fock (MCDF) method \cite{itano2}. In contrast to our RCC approach, these 
methods account less correlation effects at the same level of excitations.
\begin{table}
\caption{Comparison of $Q_{nuc}$ values (in $b$) from various works.}
\begin{ruledtabular}
\begin{center}
\begin{tabular}{lc}
 $Q_{nuc}$  &  Reference\\ 
\hline
   & \\
 $-$0.0444(6) & This work \\
 $-$0.040(8) & \cite{arnold} \\
 $-$0.0408(8) & \cite{sundholm} \\
 $-$0.043(9) & \cite{silverans} \\
 $-$0.065(20) & \cite{grundevik} \\
 $-$0.049(5) & \cite{salomonson,olsson} \\
 $-$0.044 & \cite{yu} \\
 $-$0.062(12) & \cite{aydin} \\
\end{tabular}
\end{center}
\end{ruledtabular}
\label{tab5}
\end{table}
\begin{table}
\caption{Calculations of $B_{hfs}$ (in MHz) using new $Q_{nuc}$ value and our  
$B_{hfs}/Q_{nuc}$ results reported in Table \ref{tab1}.}
\begin{ruledtabular}
\begin{center}
\begin{tabular}{lc}
 State   & $B_{hfs}$  \\ 
\hline
   & \\
4p $^2P_{3/2}$ & $-$6.715(125) \\
5p $^2P_{3/2}$ &  $-$2.260(56) \\
3d $^2D_{3/2}$ & $-$2.987(71) \\
4d $^2D_{3/2}$ & $-$0.774(16) \\
3d $^2D_{5/2}$ & $-$4.239(115) \\
4d $^2D_{5/2}$ & $-$1.100(24) \\
\end{tabular}
\end{center}
\end{ruledtabular}
\label{tab6}
\end{table}

\begin{table*}
\caption{Hyperfine interaction energies ($W_F$) for different $|F;I,J\rangle$ states in $^{43}$Ca$^+$ in MHz.}
\label{tab7}
\begin{tabular}{l|cccccccccc}
\hline
\hline
\backslashbox{$F$}{$|JM_J\rangle$} & 4s $^2S_{1/2}$ & 5s $^2S_{1/2}$ & 4p $^2P_{1/2}$& 5p $^2P_{1/2}$ & 4p $^2P_{3/2}$ & 5p $^2P_{3/2}$ & 3d $^2P_{3/2}$ & 4d $^2P_{3/2}$ & 3d $^2P_{5/2}$ & 4d $^2D_{5/2}$ \\
\hline
 & & & \\
 1 &  &  &  &  &  &  &  &  & 38.477 & 32.834 \\
 2 &  &  &  &  & 201.400 & 68.551 & 317.553 & 63.535 & 32.444 & 27.207 \\
 3 & 1814.371 & 526.500 & 327.201 & 111.715 & 115.087 & 39.160 & 177.841 & 35.666 & 22.940 & 18.647 \\
 4 & $-$1411.177 & $-$409.500 & $-$254.489 & $-$86.889 & $-$4.475 & $-$1.534 & $-$10.434 & $-$2.009  & 9.421 & 7.015 \\
 5 &  &  &  &  & $-$161.121 & $-$54.824 & $-$248.977 & $-$49.932 & $-$8.840 & $-$7.880 \\
 6 &  &  &  &  &  &  &  &  & $-$32.752 & $-$26.271 \\
\hline
\hline
\end{tabular}
\end{table*}
As seen in Tables \ref{tab3} and \ref{tab4}, the trend of correlation effects
in $A_{hfs}$ calculations in the first five low-lying states are same as
discussed in our previous works \cite{sahoo2,sahoo6} except the differences
in the magnitudes are due to the new basis functions. The amount of correlation 
effects in the higher excited states are comparatively smaller. Likewise in
$A_{hfs}$, both the pair correlation and core-polarization effects which
arise through $S_{1v}$ and $S_{2v}$ RCC operators \cite{sahoo2,sahoo6},
respectively, play major roles in obtaining the final results of $B_{hfs}/Q_{nuc}$.
The differences between our $B_{hfs}/Q_{nuc}$ result of the 3d $^2D_{5/2}$
state with others is due to the accurate treatment of these correlation effects
in the present work. Combining our $B_{hfs}/Q_{nuc}$ result with the measured
$B_{hfs}$ value of this state \cite{benhelm2}, we get $Q_{nuc}=-0.0444(6)b$.
Following the same procedure when we combine $B_{hfs}/Q_{nuc}$ results of the
4p $^2P_{3/2}$ and 3d $^2D_{3/2}$ states with their corresponding experimental
$B_{hfs}$ results, it gives less accurate values as $Q_{nuc}=-0.044(10)b$ and 
$Q_{nuc}=-0.059(29)b$, respectively. The associated larger errors are mainly
due to larger uncertainties in the experimental results. We compare the above 
accurately estimated $Q_{nuc}$ value
with the previously reported results in Table \ref{tab5}. Our result is around
10\% improvement over the considered standard value $-0.049(5)b$ \cite{ragh} 
in this system. Recently Yu et al. \cite{yu} had evaluated this value as 
$-0.044b$, which is compatible with our result, by combining their 
$B_{hfs}/Q_{nuc}$ result of the 4p $^2P_{3/2}$ state with its experimental
$B_{hfs}$ value which has around 20\% uncertainty. Sundholm and Olsen
\cite{sundholm} had combined precisely measured $B_{hfs}$ of the 3d 4s $^1D_2$
state of $^{43}$Ca with their calculated electric field gradient result
to obtain $Q_{nuc}=-0.0408(8)b$ in the same atom. In this work,
they had employed the MCDF method on the restricted active space to calculate
electric field gradient. We have already discussed the difference between the
RCC and MCDF methods earlier in this section. There are also other works
\cite{silverans,salomonson,olsson,arnold,aydin,grundevik} finding $Q_{nuc}$ 
in $^{43}$Ca, but all of them have used either large uncertainty experimental 
results or more approximated theoretical methods like second order many-body 
perturbation theory, results correcting for Sternheimer effects using the
anti-shielding factor in the Hartree-Fock calculations etc. that cope with
lesser electron correlation effects than our RCC method.

Using our new $Q_{nuc}$ value, we determine $B_{hfs}$ from our $B_{hfs}/Q_{nuc}$
results presented in Table \ref{tab1} and have given them in Table \ref{tab6}.
The new $B_{hfs}$ results are well within the error bars of experimental values with less 
uncertainties. Again, we evaluate energies of different hyperfine states using 
the formula given by Eq. (\ref{eqn2}) corresponding to each fine structure 
level. In Table \ref{tab7}, we have reported these results which can be 
verified by analyzing isotope shift measurements in the future experiments 
 in the assumed system. 

\section{Conclusion}
We have employed the relativistic coupled-cluster method in the Coulomb and Breit
interaction approximation to calculate the atomic wave functions in $^{43}$Ca$^+$. Using these wave functions, we were able to determine $A_{hfs}$ and $B_{hfs}/Q_{nuc}$ results accurately. By combining our $B_{hfs}/Q_{nuc}$ result of the 
3d $^2D_{5/2}$ state with its corresponding precisely measured $B_{hfs}$
value, we determine $Q_{nuc}$ of $^{43}$Ca as $-0.0444(6)b$ which is 10\%
improvement over the considered standard value. In this work, we have also
given contributions separately from the Breit interaction and hyperfine 
interaction energies for a number of states. Our new $Q_{nuc}$ value and 
reported hyperfine interaction energies may serve the researcher of both 
atomic and molecular physics.

\section{Acknowledgment}
This work is supported by NWO under VENI fellowship grant with project number
680-47-128. We are indebted to B. P. Das for encouraging us to carry out this 
work. We also gratefully acknowledge discussions with C. Roos. The calculations 
were carried out using the Tera-flopp Super computer in C-DAC, Bangalore.

\end{document}